\documentclass[10pt, conference, letterpaper]{IEEEtran}
\usepackage{algorithmicx}
\usepackage[ruled,vlined,linesnumbered]{algorithm2e}
\usepackage{hhline}
\usepackage{amsmath,mathtools}
\usepackage{amsfonts,amssymb}
\usepackage{mathrsfs}
\usepackage{gensymb} 
\usepackage{caption}
\usepackage{enumitem}
\usepackage{multirow}
\usepackage{graphicx} 
\usepackage{multirow}
\usepackage{enumitem,color}
\usepackage{algpseudocode}
\begin{document}
%
\title{\huge \emph{LiveMap}: Real-Time Dynamic Map in Automotive Edge Computing \vspace{-0.1in}}

\author{\IEEEauthorblockN{Qiang Liu, Tao Han, Jiang (Linda) Xie \vspace{-0.15in}}\\
\IEEEauthorblockA{The University of North Carolina at Charlotte\\
\{qliu12, tao.han, linda.xie\}@uncc.edu}\vspace{-0.3in}
\and
\IEEEauthorblockN{BaekGyu Kim \vspace{-0.15in}}\\
\IEEEauthorblockA{Toyota Motor North America R\&D InfoTech Labs \\
baekgyu.kim@toyota.com}\vspace{-0.3in}
}

\maketitle

\begin{abstract}
Autonomous driving needs various line-of-sight sensors to perceive surroundings that could be impaired under diverse environment uncertainties such as visual occlusion and extreme weather.
To improve driving safety, we explore to wirelessly share perception information among connected vehicles within automotive edge computing networks.
Sharing massive perception data in real time, however, is challenging under dynamic networking conditions and varying computation workloads.
In this paper, we propose \emph{LiveMap}, a real-time dynamic map, that detects, matches, and tracks objects on the road with crowdsourcing data from connected vehicles in sub-second.
We develop the data plane of \emph{LiveMap} that efficiently processes individual vehicle data with object detection, projection, feature extraction, object matching, and effectively integrates objects from multiple vehicles with object combination.
We design the control plane of \emph{LiveMap} that allows adaptive offloading of vehicle computations, and develop an intelligent vehicle scheduling and offloading algorithm to reduce the offloading latency of vehicles based on deep reinforcement learning (DRL) techniques.
We implement \emph{LiveMap} on a small-scale testbed and develop a large-scale network simulator.
We evaluate the performance of \emph{LiveMap} with both experiments and simulations, and the results show \emph{LiveMap} reduces 34.1\% average latency than the baseline solution.
\end{abstract}


\section{Introduction}
\label{sec:introduction}

As the development of contemporary artificial intelligence and parallel computing hardware, autonomous driving and advanced driving assistance system (ADAS) are becoming a reality more than ever~\cite{liu2019edge}.
Vehicles rely on sensors such as stereo camera, LiDAR and radar, to perceive surrounding environments, and depend on advanced onboard processors to process the massive volume perception data in real time.
By understanding the environment, e.g., high-accurate localization, lane detection and pedestrian recognition, alongside the accurate high-definition (HD) map, intelligent control algorithms can correctly react to most of the environmental situations by controlling the vehicle, e.g., lane changing and passing vehicles.

It is, however, very difficult to realize high reliable and safe driving under extremely diverse environmental uncertainties such as extreme weather, lighting conditions, visual occlusion, and sensor failures~\cite{yaqoob2019autonomous}.
For example, vehicle sensors are primarily range-limited and line-of-sight, which means they are incapable of perceiving information from occluded areas~\cite{qiu2018avr}.
Consider a single-lane road, a car is following a big truck that occludes the car's sensor perception, passing the truck without sufficient information about the opposite lane is absolutely unsafe.
Besides, the perception range and fidelity of sensors might be further impaired by extreme weather, e.g., rain, snow, and dust.
Thus, relying on sensors in a single vehicle alone may not be sufficient to fulfill high-safety driving.

Connected vehicle is the key building block of the Internet of Vehicles (IoV)~\cite{contreras2017internet}, which connects vehicles with wireless technologies, e.g., cellular networks and dedicated short-range communications (DSRC).
It allows the communication of vehicle-to-vehicle (V2V), vehicle-to-infrastructure (V2I), and vehicle-to-network (V2N), and could substantially improve driving safety by effectively sharing information among vehicles.
For example, if the truck shares its perception data to the car, e.g., a detected vehicle on the opposite lane, the car can decide not to pass the truck even if the vehicle is unobserved by the car's sensors.

\begin{figure}[!t]
	\centering
	\includegraphics[width=3.0in]{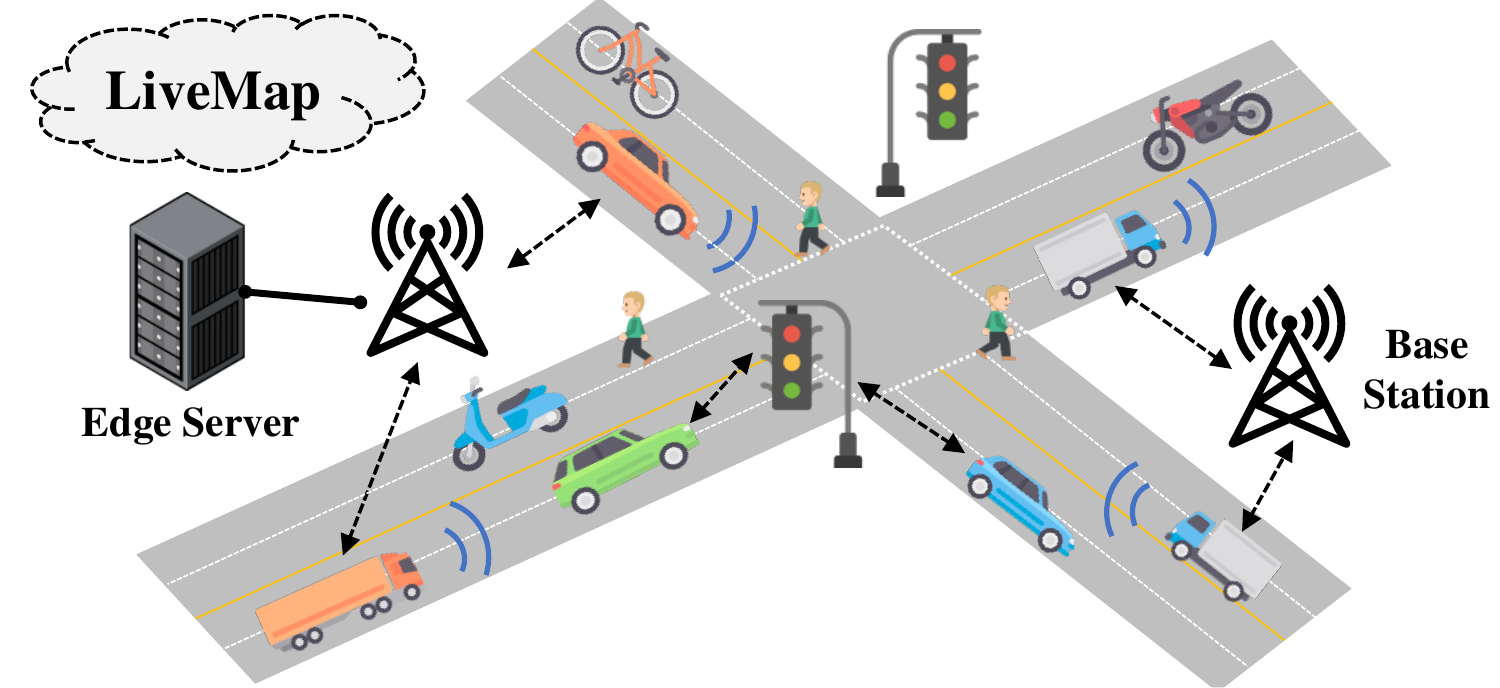}
	\vspace{-0.05in} \caption{\small An illustration of automotive edge computing.}
	\label{fig:example}
\end{figure}

However, sharing perception data among connected vehicles in automotive edge computing networks is challenging.
For example, in a dense urban scenario, allowing all vehicles to share could lead to redundant information exchanging since their sensing coverages are heavily overlapped.
Besides, letting vehicles to share their raw data, e.g., point clouds or RGB-D images, requires tremendous wireless bandwidth and might result in network congestion~\cite{ahmad2020carmap}.
Furthermore, edge servers need to support hundreds or thousands of vehicles simultaneously, thus its workloads vary from time to time.
Optimizing the time to share for vehicles needs intelligence under the complex networking and computation in automotive edge computing networks.

\begin{figure*}[!t]
	\centering
	\includegraphics[width=6.8in]{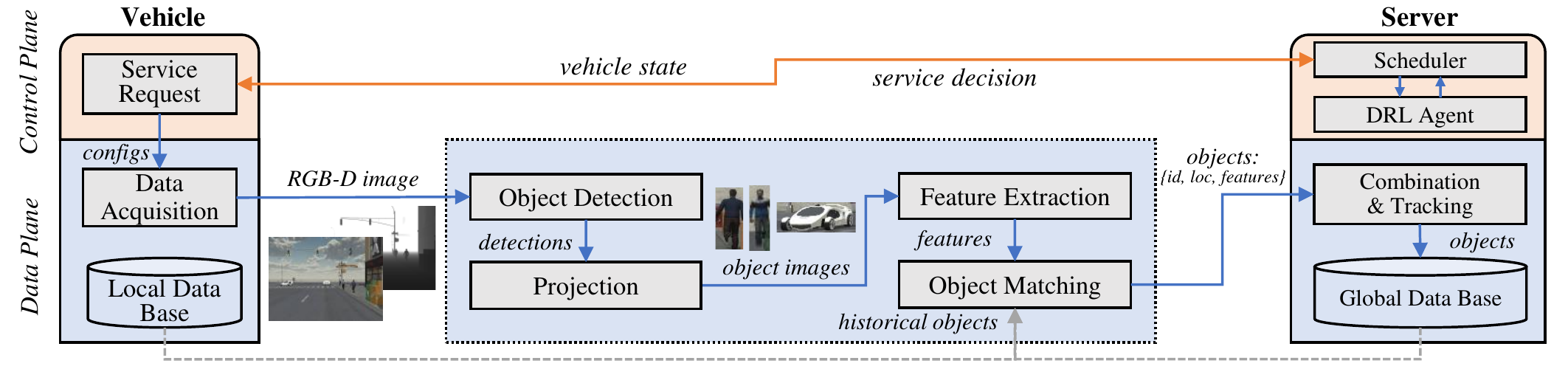}
	\vspace{-0.05in}
	\caption{\small The \emph{LiveMap} system}
	\label{fig:architecture}
\end{figure*}

In this paper, we propose \emph{LiveMap}, a real-time dynamic map in automotive edge computing illustrated in Fig.~\ref{fig:example}, that detects, matches and tracks objects on the road in sub-second based on the crowdsourcing data from connected vehicles.
We design the data plane of \emph{LiveMap} that consists of object detection, projection, feature extraction and object matching for processing individual vehicle data efficiently, and object combination for combining objects from multiple vehicles effectively.
Specifically, we reduce the detection time with neural network pruning techniques in the object detection, decrease the object feature size with variational autoencoder in the feature extraction, and improve the matching accuracy with a novel location-aware distance function.
To efficiently exploit available networking and computation resources, we design the control plane of \emph{LiveMap} by allowing adaptive offloading of vehicle computations, and developing an intelligent vehicle scheduling and offloading algorithm to decrease the offloading latency of vehicles while maintaining the map coverage based on deep reinforcement learning approaches.



The contributions of this paper are summarized as follows:
\begin{itemize}
    \item We design \emph{LiveMap} that realizes real-time dynamic map with crowdsourcing vehicles in automotive edge computing networks. 
    \item We develop the data plane of \emph{LiveMap} with an efficient processing pipeline for processing individual vehicle data and an effective object combination method for combining objects from multiple vehicles.
    \item We design an intelligent vehicle scheduling and offloading algorithm that significantly reduces the offloading latency of vehicles in the control plane of \emph{LiveMap}.
    \item We implement \emph{LiveMap} on a small-scale testbed and develop a large-scale network simulator.
    \item We evaluate \emph{LiveMap} with both experiments and simulations, and the results validate \emph{LiveMap} substantially outperforms existing solutions.
\end{itemize}

\section{\emph{LiveMap} Overview}
\label{sec:system_overview}
The control plane and data plane of \emph{LiveMap} are shown in Fig.~\ref{fig:architecture}. 
The control plane is composed of a service request module on individual vehicle, a scheduler and a DRL agent on the edge server.
When a vehicle attempts to start the offloading service, it sends a service request with current vehicle states such as computation capability and wireless quality to the scheduler.
The scheduler determines if this vehicle is scheduled based on current map coverage, and the DRL agent will be engaged to optimize the adaptive offloading decision under the current vehicle and system state if the vehicle is scheduled.
Once the service decision is received by the vehicle, it starts the data plane accordingly if scheduled, otherwise, it waits for a backoff time before next requesting. 

The data plane consists of multiple sequential processing modules for vehicle raw data processing, i.e., data acquisition, object detection, projection, feature extraction, object matching and combination.
Depending on the adaptive offloading decision of the vehicle, these modules are executed on either vehicle or server, where the data acquisition and combination module can only be run on the vehicle and server, respectively.
The data acquisition module captures RGB-D images from onboard sensors and obtains the localization of the vehicle.
The object detection module detects objects in the RGB image with a state-of-the-art object detection framework, i.e., YOLOv3~\cite{redmon2018yolov3}.
These detected objects that are with 2D bounding boxes, are projected to 3D world locations with the depth image and camera-to-world matrix obtained in the data acquisition module.
The feature extraction module extracts the features of detected objects from cropped images based on a novel variational autoencoder.
Then, these objects are matched within either local or global database depending on where to execute the matching module, based on their features and locations.
The combination module on the server integrates duplicated objects from multiple vehicles by combining the detection confidence, geo-location, and features of objects, and also predicts their mobility.
Finally, the global database is updated, whose new updates, e.g., new objects or updated locations, are broadcasted to all connected vehicles to update their local databases.

\vspace{-0.05in}
\section{Data Plane Design}
\label{sec:data_plane}
The design of data plane is to accomplish the data processing of vehicle raw data and to optimize the performance in terms of accuracy and processing delay. 
\vspace{-0.05in}
\subsection{Data Acquisition}
The data acquisition module on the vehicle is used for acquiring the raw vehicle data, i.e., RGB-D images, from various sensors, e.g., front and rear cameras or LiDARs.
In addition, it obtains the current accurate 3D world location of the vehicle based on either advanced relocalization algorithms, e.g., ORB SLAM2~\cite{murORB2}, or high-accuracy GPS, which could achieve centimeter-level estimation errors of vehicle locations.
Since we need to combine the detected objects from crowdsourcing vehicles on the edge server, the accurate estimation of vehicle locations are desired to improve the system performance.

\vspace{-0.05in}
\subsection{Object Detection}
The object detection module detects objects in the RGB image, e.g., pedestrians, cars and trucks, where detection results are composed of object names, confidences and 2D bounding boxes.
Existing object detection algorithms, e.g., YOLO~\cite{redmon2018yolov3}, Fast-RCNN~\cite{girshick2015fast}, SSD~\cite{liu2016ssd}, are primarily designed for detecting generic categories which include hundreds of object classes, e.g., person, book, boat, table and kite.
As a result, applying these algorithms directly to embedded platforms such as electronic control unit (ECU) in vehicles, leads to long detection time.
In general, a larger neural network is required to achieve similar detection accuracy, e.g., mean average precision (mAP), for a larger number of classes.

\begin{figure}[!t]
	\centering
	\includegraphics[width=3.1in]{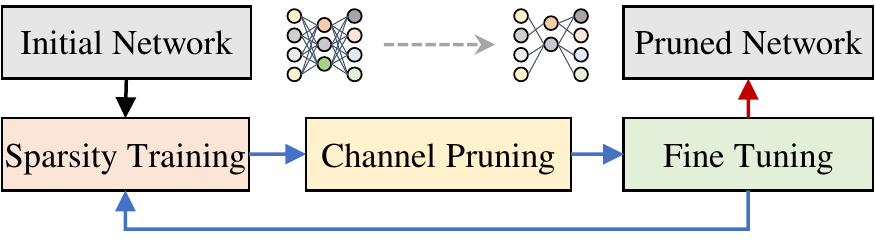}
	\vspace{-0.05in} \caption{\small The flow-chat of neural network pruning.}
	\label{fig:network_pruning}
\end{figure}

To address this issue, we propose a slim object detector with reduced categories specifically for automotive transportation system, by using neural network pruning techniques.
The neural network pruning is to reduce the neural network size by removing unnecessary neurons without dramatically scarifying detection accuracy.
As shown in Fig.~\ref{fig:network_pruning}, we adopt a similar network pruning workflow in \cite{Liu_2017_ICCV}, which mainly consists of sparsity training, channel pruning, fine-tuning.
Specifically, the initial network is re-trained by minimizing the loss function with a weighted L1 regulation on the scaling factors in batch normalization (BN) layers during the sparsity training phase.
By decreasing these BN scaling factors, insignificant convolutional channels that are with nearly zeros scaling factors can be pruned during the channel pruning phase.
Then, the neural network is fine tuned during the fine-tuning phase.
Such train-prune-tune sequential processes can be repeated to seek the optimal trade-off between detection accuracy and network size.

As a result, we show the performance of detection network pruning in Table~\ref{tbl:nework_pruning}.
We apply the network pruning on the YOLOv3 tiny framework, where we decrease 80 classes to 10 classes that includes person, bicycle, car, motorcycle, airplane, bus, train, truck, boat, and traffic light.
We reduce the network size by 93.7\% with the cost of 0.01 mAP degradation. 
Meanwhile, the detection time on Nvidia Jetson Nano is decreased by 19.4\% or 18.7\% if using TensorRT acceleration.

\vspace{-0.05in}
\subsection{Projection}
The projection module calculates the 3D world location of detected objects in the world coordinate system based on the detection results, i.e., 2D bounding boxes of objects, depth image, and camera specifications.

\begin{figure}[!t]
	\centering
	\includegraphics[width=3.1in]{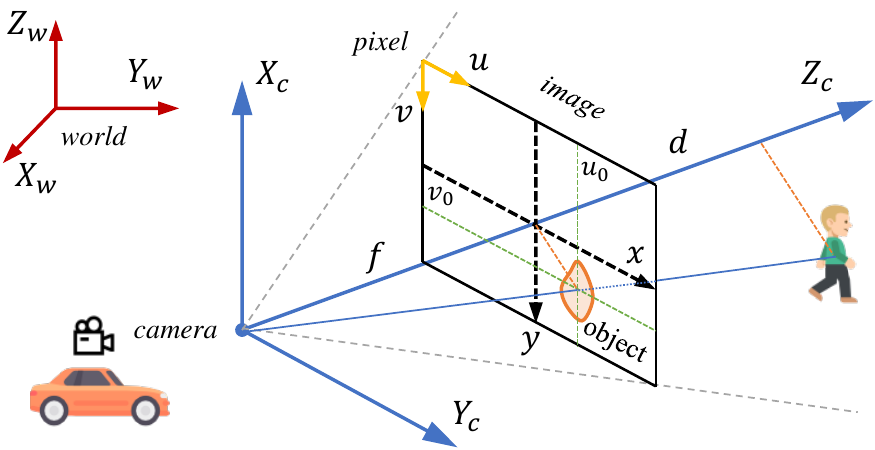}
	\vspace{-0.05in} \caption{\small The illustration of coordinate systems.}
	\label{fig:projection}
\end{figure}

As shown in Fig.~\ref{fig:projection}, the objects in real world are projected onto the image plane by the camera sensor.
The calculation of the world location of an object is completed by two steps, i.e., from pixel coordinates to camera coordinates, and from camera coordinates to world coordinates.
Denote $(u_0, v_0)$ as the center of an object in pixel coordinates, the focal length of the camera as $f$, and the image resolution as $(R_W, R_H)$. 
The 3D location of the object $(X, Y, Z)$ in the camera coordinate system can be written as
\begin{align}
    X \; &= \; - (d*(v_0 - 0.5*R_H)) / f, \nonumber \\ 
    Y \; &= \; (d*(u_0 - 0.5*R_W)) / f, \\ 
    Z \; &= \; d,  \nonumber
\end{align}
where $d$ is the Z-axis depth of the object w.r.t. the camera.
However, estimating the depth of an object is not easy since the object usually occupies an irregular 2D area in the depth image while its bounding box only gives the rectangle area.
Given the bounding box of the object and depth image, we sample multiple small 5x5 squares in close proximity to the object center and calculate the average depth after removing the largest and smallest values.

Next, we convert the object location in camera coordinate to the world location \textbf{$(W_x, W_y, W_z)$ } in world coordinate as
\begin{equation}
    \left[W_x,\; W_y,\; W_z,\;1\right]^T = M_{c2w} \times \left[X,\;Y,\;Z,\;1 \right]^T,
\end{equation}
where $M_{c2w}$ is the 4x4 camera-to-world conversion matrix obtained from the data acquisition module, and $[\cdot]^T$ is the operation of matrix transpose.

In addition, the projection module estimates the coverage of a vehicle by obtaining the default coverage of its cameras and calculating the visual occlusion incurred by objects.
In \emph{LiveMap}, we consider the area is occluded by an object if the object height is higher than that of the camera in the camera coordinate system, i.e., $X >= 0$.

\begin{table}[!t]
    \small
	\centering
	\vspace{0.1in}
	\begin{tabular}{|c||c|c|c|c|}
        \hline
        \textbf{Detection}     &  \textbf{mAP@0.5}  &  \textbf{ Num. of }  &  \textbf{time(Nano)}  \\ 
        \textbf{Networks}  &   640x  &  \textbf{parameters}   &  w/o TensorRT \\ 
        
        \hline
        YOLOv3 tiny  &     0.534                  &    8.69e+06 & 191.9/37.4 ms   \\ 
        \hline
        Pruned Network   &      0.524                  &   0.54e+06 &  154.7/30.4 ms \\ 
        \hline
	\end{tabular}
	\vspace{-0.05in} \caption{ \small Pruning results of detection network on our dataset}\label{tbl:nework_pruning}
\end{table}

\vspace{-0.05in}
\subsection{Feature Extraction}
Although we obtain the world location of all detected objects by the detection and projection modules, we need to identify and match them within the database to track their mobility.
The feature extraction module is used to extract features from cropped object images based on variational autoencoder framework.
The conventional feature extraction algorithms, e.g., SIFT, SURF~\cite{bay2008speeded}, and ORB~\cite{rublee2011orb}, could generate keypoint features with similar data size as compared to that of the object image~\cite{zhang2018jaguar}.
Consequently, the computation complexity of feature matching raises and the transmission delay of features offloading if applicable is increased accordingly.
Besides, the detected objects are usually small, e.g., pedestrian images could be 50x50 out of 540p camera images, because they are tens of meters away from vehicles.
In practice, we found that these algorithms either generate no features or trivial features from small objects, which results in low matching accuracy.

\begin{figure}[!t]
	\centering
	\includegraphics[width=3.4in]{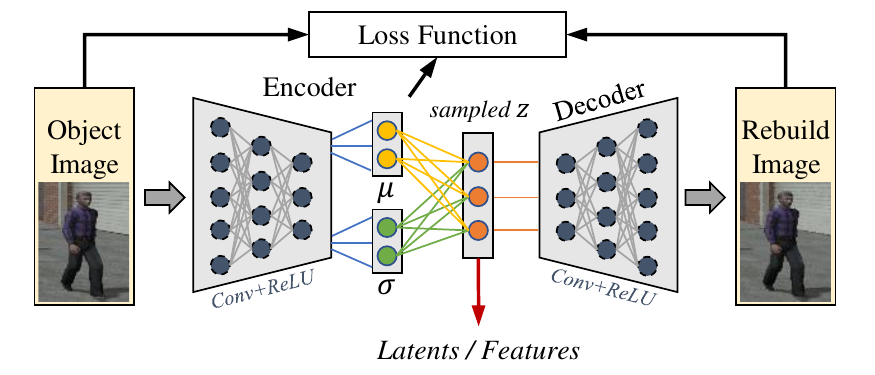}
	\vspace{-0.05in} \caption{\small The design of autoencoder as feature extraction.}
	\label{fig:autoencoder}
\end{figure}

To solve this issue, we propose to use variational autoencoder~\cite{kingma2013auto}, an unsupervised machine learning framework, to extract lightweight features from object images, as shown in Fig.~\ref{fig:autoencoder}.
The autoencoder primarily consists of an encoder that encodes the input image into condensed latent vectors and a decoder that rebuilds the image from the latents.
Unlike conventional autoencoders, which are prone to generate irregular latent space~\cite{kingma2013auto}, e.g., similar images may be encoded to distinct latent vectors, the variational autoencoder uses a unique neural network architecture and introduces a regularization in the loss function.
Denote $x, z$ as the object image and the sampled latent vector from the distribution $\mathcal{N}(\mu, \sigma^2)$, respectively.
The training loss of variational autoencoder can be written as 
\begin{equation}
    Loss = - \mathbb{E}_{z \sim q(z|x)} \left[  \log p(x|z)  \right] + D_{KL} \left[ q(z|x) | p(z)\right], 
\end{equation}
where $q(z|x)$ and $p(x|z)$ denote the encoder and decoder, respectively. And $D_{KL}$ is the Kullback-Leibler divergence to evaluate the difference between two probability distributions, where $p(z) \sim \mathcal{N}(0,1)$ is selected as a normal distribution.
After the training of the variational autoencoder, the encoder is used to extract features from object images and the generated latent vectors are recognized as the features of objects.

\vspace{-0.05in}
\subsection{Object Matching}
The object matching module is designed to match the detected objects within the database based on their features and locations.
Since the database in \emph{LiveMap} could have hundreds or thousands of items such as pedestrians and cars, matching an object with all these items is compute-intensive and time-consuming.
Meanwhile, matching objects merely based on the distance of features, i.e., latent vectors, could fail in a dynamic automotive environment~\cite{ahmad2020carmap}.

To solve this issue, we propose a novel location-aware distance function for matching based on our estimated world locations of vehicles and objects.
Specifically, we only match the database items in close proximity to the detected object, e.g., 100 meters for vehicles and 10 meters for pedestrians, which decreases the size of matching set and reduces the matching time accordingly.
In \emph{LiveMap}, we construct a mobility model for each object in the database based on its historical locations.
The location of objects in the database are predicted when matching objects at the current time.
Then, we introduce a novel location-aware distance function by considering not only the features distance but also the geographic distance between two objects.
Denote $g$ as the geo-location of an object in the world coordinate system, the distance between the $i$th and $j$th object is defined as 
\begin{equation}
    D_{i,j} = \min(\left[||z_{i,m}-z_{j,m} ||^2, \forall m \in \mathcal{M}\right]) + w ||g_i -g_j||^2, \label{eq:matching}
\end{equation}
where $w$ is a weighted factor, $z$ are the latent features of objects, $||\cdot||^2$ is the L2-norm operation, and $\mathcal{M}$ denotes the set of latent features associated with an item in the database.
Since an object may be observed by multiple vehicles from different angles, these multi-view generated features are associated with the object in the database.
Here, we use the minimum feature distance among these multi-view features to calculate the final distance between two objects. 

\begin{table}[!t]
    \small
	\centering
	\vspace{0.1in}
	\begin{tabular}{|c|c|c|c|}
        \hline
        object id    &  class    &    geo-location &  confidence       \\ 
        \hline
        speed      &    direction &  update time      &     multi-view latents     \\ 
        \hline
	\end{tabular}
	\vspace{-0.05in} \caption{ Attributes of an object in the database }\label{tbl:database_item}
\end{table}

\vspace{-0.05in}
\subsection{Object Combination}
The object combination module on the server integrates the detected objects from different vehicles and updates their information in the global database, e.g., locations and latent features.
The global database is a collection of historical detected objects, where each object is represented by several attributes as shown in Table~\ref{tbl:database_item}. In \emph{LiveMap}, we remove objects from the global database if their information is outdated, e.g., a pedestrian is deleted if not observed for more than 1 hour.

Due to the high diversity of vehicles, e.g., camera specs, view angles and lighting, an object captured and processed by different vehicles may generate slightly different results in terms of confidence, location and latents features.
To effectively integrate these results together, we first consider the objects with the same object id as a unique object, and then propose a confidence weighted combination method that calculates the geo-location of the unique object as
\begin{equation}
    g =  \sum\limits_{m \in \mathcal{M}} \frac{ \mathcal{P}_m * g_m }{\sum\limits_{m \in \mathcal{M}} \mathcal{P}_m } ,
\end{equation}
where $\mathcal{P}_m$ is the confidence generated by the object detection module, and $g_m$ is the geo-location estimated by the projection module.
Here, we assign more weights to the results with higher confidence.
Meanwhile, we consider each latent feature of a unique object is valid and associate it with the object into the database for better matching accuracy as shown in Eq.~\ref{eq:matching}.

Finally, these unique objects are updated and stored in the global database.
The new updates at the current time in the global database, e.g., newly detected objects, new location and latents of existing objects, are broadcasted to all connected vehicles in \emph{LiveMap}.

\vspace{-0.05in}
\section{ Control Plane Design}
In this section, we describe the system model, formulate the vehicle scheduling and offloading problem in \emph{LiveMap}, and develop a novel algorithm to solve the problem efficiently.

\vspace{-0.05in}
\subsection{System Model}
We consider an automotive edge computing network with multiple vehicles denoted as $\mathcal{I}$, a cellular base stations (BS) and an edge computing server, where vehicles are wirelessly connected to the BS.
Connected vehicles offload vehicle computations to the edge server asynchronously to build \emph{LiveMap}.
We consider vehicle computations, e.g., the data plane in \emph{LiveMap}, can be separated between computation modules\footnote{The discrete separation model can be easily extended for different systems, such as partial neural network offloading in AR/VR system~\cite{ran2017delivering}.}, denoted $y_i \in \{0, 1, ..., N\}, \forall i \in \mathcal{I}$, where $N$ is the maximum separation scheme.
For example, if the separation scheme is $1$ in \emph{LiveMap}, it means the object detection module is executed on the vehicle and the remaining modules, i.e., projection, feature extraction and object matching, are processed on the server.
Meanwhile, the intermediate data generated by the object detection module, i.e., detection results and cropped object RGB-D images, are offloaded to the edge server for remaining processing.

Before the computation offloading, a connected vehicle sends an offloading request to the scheduler in the edge server with some vehicle information, e.g., wireless quality and computation capacity.
Denote $x_i \in \{0, 1 \}, \forall i \in \mathcal{I}$ as the binary schedule indicator of the $i$th vehicle, where $x_i=1$ means the current request of the vehicle is scheduled, otherwise, the request is not scheduled.
Denote $C_i^{(t)}$ as the geographic coverage area of the $i$th vehicle at the $t$ time slot.
Denote $\mathcal{X, Y}$ as the set of vehicle scheduling and offloading decision for all vehicles, respectively.
Define the latency $L_i^{(t)}$ of the $i$th vehicle at the $t$ time slot as the time between the vehicle gets the raw data by the data acquisition module and the vehicle receives the broadcasted database updates from the edge server.





\subsection{Problem Formulation}
On maintaining \emph{LiveMap}, our objective is to provide information about the environment to all connected vehicles as fast as possible.
Due to the high mobility of vehicles, the outdated information is less meaningful, e.g., a recorded location of a truck 10 seconds ago does not help on making controlling decisions in a highway scenario.
Meanwhile, \emph{LiveMap} should maintain sufficient coverage areas by scheduling more crowdsourcing vehicles where each vehicle covers a certain area on its current location.
Here, we define the overall map coverage at the current time as $\textstyle \bigcup\limits_{i\in \mathcal{I}} C_i^{(t)}$.

Therefore, we formulate the vehicle scheduling and offloading problem as
\begin{align}
\centering
\label{prob0}
&{\min \limits_{ \{\mathcal{X,Y}\}} \;\;\;\;\;\;  \sum\limits_{t \in \mathcal{T}}{\sum\limits_{i \in\mathcal{I}}L_i^{(t)}}} \\ 
&{\;\;s.t.\;\;\;\;\;\;\textstyle \bigcup\limits_{i\in \mathcal{I}, x_i \neq 0} C_i^{(t)} \ge \beta \textstyle \bigcup\limits_{i\in \mathcal{I}} C_i^{(t)}, \forall t \in \mathcal{T}}, \label{prob:const1} \\ 
&{\;\;\;\;\;\;\;\;\;\;\;\;\;x_i^{(t)} \in \{0, 1\}, \forall i \in \mathcal{I}},  t \in \mathcal{T}, \label{prob:const2} \\ 
&{\;\;\;\;\;\;\;\;\;\;\;\;\;y_i^{(t)} \in \{0, 1, ..., N\}, \forall i \in \mathcal{I}, t \in \mathcal{T}}, \label{prob:const3}
\end{align}
where $\mathcal{T}$ is a given time period such as 1 hour, constraints in Eq.~\ref{prob:const1} guarantee the minimum requirement of overall map coverage, and $\beta \in [0, 1]$ is a factor.

The key difficulties in solving the above problem are highlighted.
First, due to the heterogeneity of vehicles in terms of computation capability and varying wireless quality, alongside the complicated of networking and computation in \emph{LiveMap}, the accurate modeling of vehicle latencies are impractical to be obtained in real systems.
Second, with the asynchronous offloading of crowdsourcing vehicles, their wireless transmissions and server computations are probably overlapped in time. 
As a result, the vehicle scheduling and offloading in \emph{LiveMap} exhibits \emph{Markovian} property on serving these connected vehicles, which further complicates the problem.


\subsection{Algorithm Design}
To effectively solve the problem, we develop a novel algorithm based on deep reinforcement learning.
The DRL techniques have shown promising improvement in network management and control ~\cite{liu2020edgeslice,wang2019intelligent} in terms of system performance, however, it is challenging to apply DRL in solving the aforementioned problem directly.
On one hand, the number of connected vehicles in \emph{LiveMap} is varying from time to time because the high-speed vehicles might come and leave the coverage of the BS. 
Most DRL solutions are designed to solve problems with fixed action space, and thus they are unable to handle the dynamic vehicle scheduling in \emph{LiveMap}. 
On the other hand, existing DRL solutions are inefficient to optimize problems with multiple constraints\footnote{Although there are some works~\cite{achiam2017constrained, tessler2018reward} target to solve constrained reinforcement learning problems, they are unable to guarantee these constraints are met at any time slots.}, i.e., the requirement of map coverage in Eq.~\ref{prob:const1}.

We address the problem by optimizing the vehicle scheduling and offloading decision in different time scales.
This is based on the observation that the offloading of vehicles run in sub-second time scale, such as vehicle latencies are hundreds of milliseconds, whereas the scheduling of vehicles can operate at second time scale.
Therefore, we design a two-layer ve\underline{H}icle sch\underline{E}duling and offlo\underline{A}ding \underline{D}ecision (HEAD) algorithm (see Alg.~\ref{alg:proposed}) in \emph{LiveMap}.
In the upper layer, we schedule the minimum number of vehicles while maintaining the requirement of map coverage.
Here, minimizing the number of scheduled vehicles corresponds to decreasing the number of offloading vehicles that share the common networking and computation resources in \emph{LiveMap}.
In the lower layer, we optimize the offloading decision for every single incoming vehicle with DRL techniques, where the action space becomes fixed.

\subsubsection{Vehicle Scheduling}
We build a complete graph $(V, E)$ where $V$ is the set of vertices that correspond to all vehicles, and $E$ is the set of edges between vertices.
Then, we define the overlapping ratio between the coverage of $i$th and $j$th vehicle as
\begin{equation}
    o_{i,j} =  \frac{C_i \bigcap C_j}{C_i \bigcup C_j},
\end{equation}
and assign $o_{i,j}$ to the edge value between the $i$th and $j$th vehicle, where $o_{i,j}=o_{j,i}$. Denote the average overlapping ratio of the $i$th vehicle as 
\begin{equation}
    O_i = \frac{1}{|\mathcal{I}|}\sum\limits_{j \in \mathcal{I}, j \neq i} o_{i,j}.
\end{equation}

Then, we greedily prune the graph $(V, E)$ by removing the $i$th vehicle if it has the largest average overlapping ratio, i.e., $i = \arg\max\limits_{k \in \mathcal{I}}O_k$.
The basic idea behind this pruning is that we continuously remove a vehicle with the minimum decrease in the overall map coverage.
The pruning processes stop until we reach the required map coverage by evaluating Eq.~\ref{prob:const1}.

\subsubsection{Offloading Decision}
To determine the offloading decision of a vehicle, we resort to the deep reinforcement learning, e.g., deep Q network~\cite{schaul2015prioritized}, that is capable of handling the complex offloading in \emph{LiveMap}.
Consider a generic reinforcement learning setting where an agent interacts with an environment in discrete decision epochs.
At each decision epoch $t$, the agent observes a state $\mathbf{s}_{t}$, takes an action $\mathbf{a}_t$, i.e., offloading decision, based on its policy $\pi_\theta$ that parameterized by neural networks with parameters $\theta$.
Then, the agent receives a reward $r(\mathbf{s}_t, \mathbf{a}_t)$, and the environment transits to the next state $\mathbf{s}_{t+1}$ according to the action taken by the agent.
The objective is to seek a optimal policy $\pi_\theta^*$ that maximizes the discounted cumulative reward $R_0 = \sum\nolimits_{t=0}^{\infty} \gamma^t r(\mathbf{s}_t, \mathbf{a}_t)$. Here, $\gamma \in [0, 1)$ is a discounting factor and the transition $\tau = (\mathbf{s}_t, \mathbf{a}_t, r_t, \mathbf{s}_{t+1})$.

Then, we define the state space, action space and reward.

\textbf{State Space}: The state space determines what information can be observed from the system by the DRL agent. The design of state space is to represent the status of \emph{LiveMap} completely and informatively. Thus, we build the state space $\mathbf{s}_{t}  \triangleq [\mathbf{s}_{t}^v,\; \mathbf{s}_{t}^s,\; \mathbf{s}_{t}^w]$, where $\mathbf{s}_{t}^v$ is \emph{vehicle status}, $\mathbf{s}_{t}^s$ is \emph{server status}, and $\mathbf{s}_{t}^w$ is \emph{system workload}. The vehicle status provides useful information about the vehicle, including wireless quality (measured by received signal strength) and computation capability (represented by the number of CPUs, CPU frequency, memory size, GPU cores and GPU frequency). The server status includes the computation capability of the edge server and the wireless bandwidth. The system workload includes the number of total connected vehicles in \emph{LiveMap} and the number of queuing vehicles on the edge server.

\textbf{Action Space}: Based on the observed state space, the DRL agent decides which offloading decision is applied to the current vehicle, which is $\mathbf{a}_t \triangleq [ y ]$.

\textbf{Reward}: When applying the offloading decision $\mathbf{a}_t$ to the vehicle under the current state space $\mathbf{s}_t$, the DRL agent will receive a reward from \emph{LiveMap}, which is defined as the negative latency of this vehicle, i.e., $r(\mathbf{s}_t, \mathbf{a}_t) \triangleq - L^{(t)}$.
In the real system, the reward is delayed because the latency can only be obtained after the vehicle offloading is completed. During this time interval, requests from other vehicles may arrive for offloading decisions.
We resolve the issue by allowing \emph{LiveMap} to temporally store state-action pairs and report the state-action-reward pairs once available to the DRL agent.

\begin{algorithm}[!t]
	\caption{The HEAD Algorithm}\label{alg:proposed}

	\KwIn{ $\beta$, $\theta^*$, $\mathbf{s}^s_{t}$, $\mathbf{s}^w_{t}$  }
	\KwOut{$x, y$}
	  
	    $\mathbf{s}^v_{t}, i \gets$ vehicle, $/*\;accept\; vehicle*/$\;
    	$\mathbf{s}_{t} \gets [\mathbf{s}^v_{t},\; \mathbf{s}^s_{t},\; \mathbf{s}^w_{t}]$, $/*\;build\; state\; */$\;
    	\If {time to schedule}
    	{
		    $x_k \gets 1,\; \forall k \in \mathcal{I}$\;
        	\While{$True$}
        	{
        	 $k \gets \arg\max\limits_{k \in \mathcal{I}, x_k \neq 0}O_k$\;
        	 $x_k \gets 0$;
        	 
        	 \If {$\bigcup\limits_{i\in \mathcal{I}, x_i \neq 0} C_i^{(t)} \leq \beta \bigcup\limits_{i\in \mathcal{I}} C_i^{(t)}$}
        	    {
        	        $x_k \gets 1$\;
        	        \textbf{break}\;
        	    }
        	}
    	}
    	\If {$x_i == 1$ (scheduled)}
    	{
        	$y$ $\gets$ $\arg\max \limits_{\mathbf{a}_{t}}Q^*(\mathbf{s}_{t}, \mathbf{a}_{t} | \theta^*)$, $/*\;get\; action */$\;
    	}
    	\Else
    	{
    	    $y \gets -1$,  $/*\;not\; scheduled*/$\;
    	}
	
    \Return{$x, y$}\;
\end{algorithm}

%
\textbf{Policy Training}:
We use Deep Q network (DQN)~\cite{mnih2015human} with prioritized experience replay (PER)~\cite{schaul2015prioritized} to train the policy of DRL agent in \emph{LiveMap}.
Denote the value function $ Q^\pi(\mathbf{s}_t, \mathbf{a}_t)$ as the expected discounted cumulative reward if the agent starts with the state-action pair $(\mathbf{s}_t, \mathbf{a}_t)$ at decision epoch $t$ and then acts according to the policy $\pi$. Thus, the value function can be expressed as $Q^\pi(\mathbf{s}_t, \mathbf{a}_t) = \mathop{\mathbb{E}}\nolimits_{\tau \sim \pi} {\left[ R_t | \mathbf{s}_t, \mathbf{a}_t \right]}$,
where $R_t = \sum\nolimits_{k=t}^T \gamma^{(k-t)} r(\mathbf{s}_k, \mathbf{a}_k) $.
Based on the Bellman equation~\cite{bellman1966dynamic}, the optimal value function $Q^*(\mathbf{s}_t, \mathbf{a}_t)$ is
\begin{equation}
    Q^*(\mathbf{s}_t, \mathbf{a}_t) = r(\mathbf{s}_t, \mathbf{a}_t) + \gamma \max\limits_{\mathbf{a}_{t+1}}Q^*(\mathbf{s}_{t+1}, \mathbf{a}_{t+1}).
\end{equation}

To obtain the optimal policy, DQN is trained by minimizing the mean-squared Bellman error (MSBE) as follow
\begin{equation}
    Loss(\theta^Q) = \mathop{\mathbb{E}}\limits_{\tau \in \mathcal{D}}{\left[ {\left( h_t - Q(\mathbf{s}_t, \mathbf{a}_t | \theta^Q)\right)}^2\right]},
\end{equation}
where $\theta^Q$ are weights of the Q-network and $\mathcal{D}$ is a replay buffer.
$h_t$ is the target value estimated by a target network
\begin{equation}
    h_t = r(\mathbf{s}_t, \mathbf{a}_t) + \gamma \max\nolimits_{\mathbf{a}_{t+1}}Q(\mathbf{s}_{t+1}, \pi(\mathbf{s}_{t+1}|\theta^{\pi'}) | \theta^{Q'}),
\end{equation}
where $\theta^{Q'}$ are weights of the target network.
The target network has the same architecture with the Q-network and its weights $\theta^{Q'}$ are slowly updated to track that of Q-network.

In the DQN, experience transitions in replay buffer are uniformly sampled, regardless of the significance of experiences. 
Prioritized experience replay (PER)~\cite{schaul2015prioritized} improves the efficiency of DQN sampling by prioritizing experience transitions in the replay buffer.
The importance of experience transitions are measured by the absolute TD error, that is 
\begin{equation}
    p \;\; \propto \;\; |h_t - Q(\mathbf{s}_t, \mathbf{a}_t | \theta^Q)|^\alpha,
\end{equation}
where $\alpha$ is a hyper-parameter.







\section{System Implementation}
\label{sec:implementation}
In this section, we implement \emph{LiveMap} on a small-scale testbed and develop a large-scale network simulator for automotive edge computing networks.

\subsection{System Prototype}
\begin{figure}[t]
	\centering
	\includegraphics[width=3.4in]{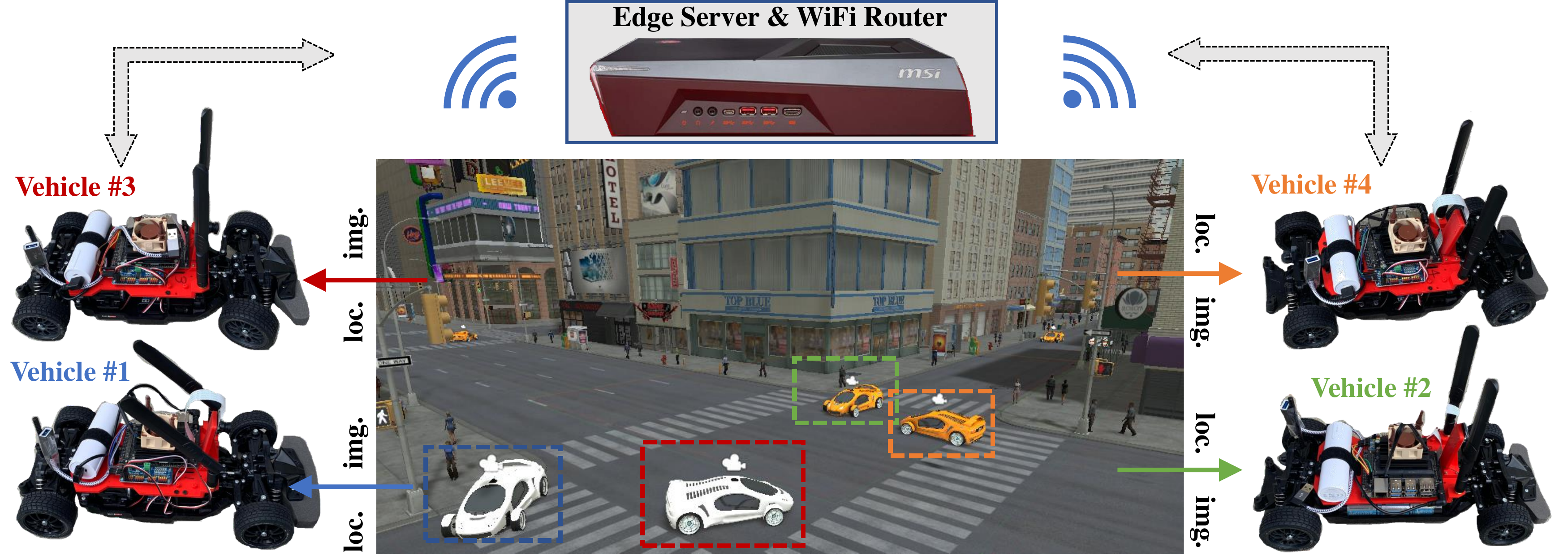}
	\vspace{-0.05in} \caption{\small The implementation of \emph{LiveMap}.}
	\label{fig:unity_env}
\end{figure}

We prototype \emph{LiveMap} system on a small-scale automotive edge computing testbed as shown in Fig~\ref{fig:unity_env}, which is composed of four JetRacers, an 802.11ac 5GHz WiFi router with 20MHz wireless bandwidth and an Intel i7 edge server with Nvidia GTX 1070 GPU and CUDA 10.1~\cite{nvidia2011nvidia}.
The JetRacer is a racecar equipped with an onboard Nvidia Jetson Nano embedded GPU.
The dynamic wireless channel of vehicles are emulated by randomly configuring the transmit power of both the JetRacers and the WiFi router with Linux "\emph{iw}" CLI configuration utility, i.e., from 1dBm to 22dBm.
On the edge server, we develop a single queue to process all incoming offloading of vehicles.
To reduce the transmission delay, we use LZ4 compression algorithm before socket communication.

We implement the DRL agent by using Python 3.7 and PyTorch 1.40.
Specifically, we use a 2-layer fully-connected neural network, i.e., [256, 256], with Leaky Recifier activiation function~\cite{goodfellow2016deep}.
The learning rate of DQN is 5e-4 with 512 batch size, and the discounted factor $\gamma=0.9$.
We add a decaying $\epsilon$-greedy starts from probability 0.5 to 0.1 during the training phase for balancing the exploitation and exploration.

Due to the Markovian property of the vehicle scheduling and offloading problem in \emph{LiveMap}, i.e., the current vehicle offloading decision could immediately affect the offloading of the next vehicle, it is ineffective to collect dataset and train the DRL agent offline, and apply the trained policy online.
Hence, we online train the DRL agent by directly interacting with JetRacers in the real \emph{LiveMap} system with 100k training steps.
To accomplish the functionalities of control and data plane in \emph{LiveMap}, e.g., object detector, autoencoder, scheduler, the DRL agent and network simulator, we finish more than 6000 line codes.

\subsection{Traces DataSet}
We build a Unity3d environment to generate traces for both testbed experiments and network simulations.
The traces are composed of more than 1000 frames, where each frame includes the world location, RGB-D images, and camera-to-world matrix of vehicles, and the world location of pedestrians for calculating estimation errors.
We create multiple transportation scenarios, e.g., intersection, highway and circle, based on the modern city package in Unity3d, where each scenario includes hundreds of pedestrians and vehicles.
The movement of pedestrians and vehicles follow their predefined paths.
Each vehicle is mounted a front RGB-D camera with focal length 50mm, field of view $54.04^{\degree}$, maximum sensing range 50m, which generates 741x540 images.

\subsection{Network Simulator}
\begin{figure}[t]
	\centering
	\includegraphics[width=3.2in]{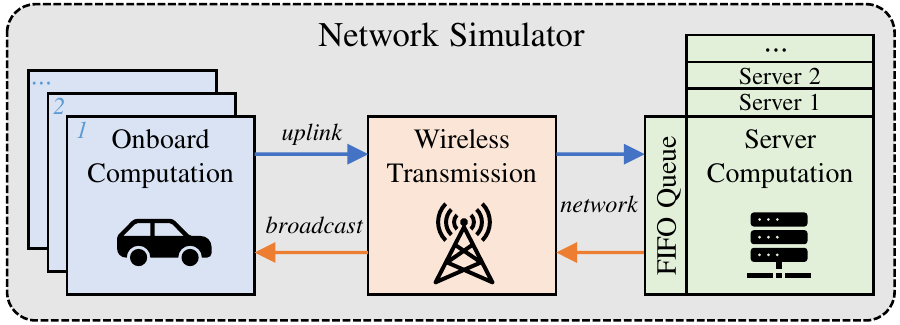}
	\vspace{-0.05in} \caption{\small The design of network simulator.}
	\label{fig:simulator}
\end{figure}

We build a time-driven network simulator, which is composed of multiple onboard computation modules, a wireless transmission module, and a server computation module as depicted in Fig.~\ref{fig:simulator}.
When the offloading decision of a vehicle is determined, a task is created on this vehicle's onboard computation module.
The task describes the remaining onboard computation time, data size of uplink and broadcast transmission, and server computation time, where these data are sampled from experimental measurements (see Fig.~\ref{fig:system_comparison}).
The onboard computation of a task is simulated by decreasing its remaining computation time for every simulation interval such as 1 ms.
The simulation of server computation is similar, but based on a single FIFO (First-In-First-Out) queue and multiple parallel servers architecture. 

The wireless transmission module is developed based on an open-source 5G simulator~\cite{oughton2019open}, where we use the urban micro (UMi - Street Canyon) channel model recommended in ETSI TR 138.901~\cite{esti_tr_138_901}, and consider all vehicles equally share the total bandwidth for the sake of simplicity.
We use both 1MHz wireless bandwidth for uplink and downlink channels, and place the base station at the center of the Unity3d environment.
Thus, the transmission of tasks are simulated by calculating their wireless data rates and decreasing their remaining uplink/downlink data sizes.
A task is sent to the next simulation module only if it is completed in the last module, e.g., zero remaining computation time or transmission data size.


\begin{figure*}[!t]
	\centering
	\includegraphics[width=6.4in]{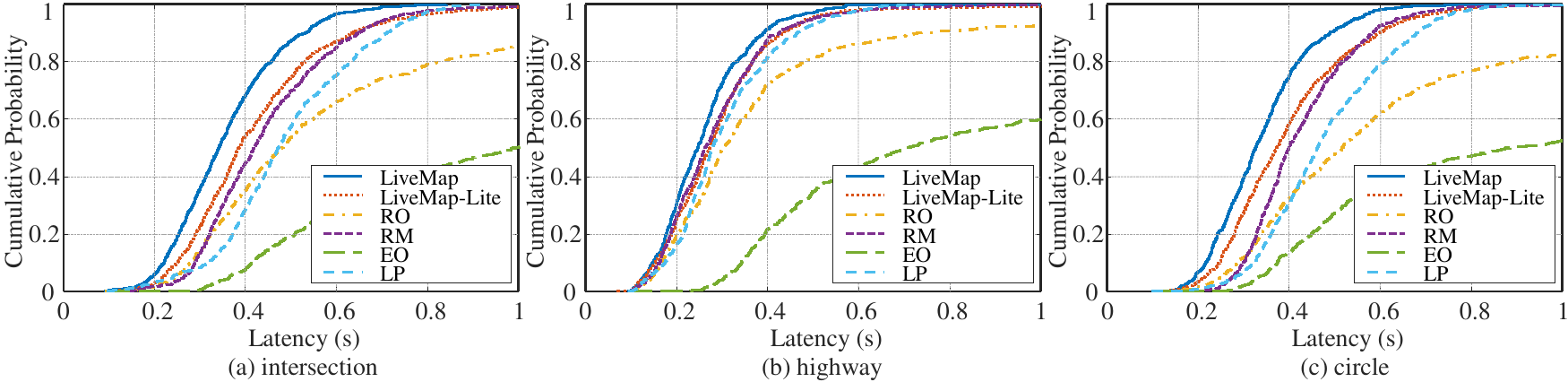}
	\vspace{-0.05in} \caption{\small The cumulative probability of latency by different algorithms under various scenarios.}
	\label{fig:scenarios}
\end{figure*}

\subsection{Comparison Algorithms}
In the experiments, we compare \emph{LiveMap} with the following algorithms:
1) Edge offloading (\textbf{EO}): EO offloads RGB-D images of vehicles and allows all the processing modules to be executed on the edge server.
2) Local process (\textbf{LP}): LP executes all the processing modules onboard, and sends the matched objects to the edge server.
3) Random offloading (\textbf{RO}): RO randomly selects the offloading decision for every vehicle.    
4) Regression model (\textbf{RM}): We propose RM that identifies wireless data rate and the number of vehicles as two important factors when making the offloading decision. Thus, RM fits a multivariate polynomial regression model with \emph{scikit-learn} tool based on an experimental dataset that includes different combinations of wireless data rate, number of vehicles, offloading decision and the resulted latency. To make the offloading decision, RM predicts the latency of different actions under the current network state, and takes the action with the minimum predicted latency.
5) \textbf{LiveMap-Lite}: LiveMap-Lite determines offloading decision as same as that of LiveMap, but it schedules all vehicle requests. 
Besides, these comparison algorithms schedule all vehicle requests.

\section{Performance Evaluation}
\label{sec:evaluation}
In this section, we evaluate the performance of \emph{LiveMap} on both small-scale testbed and large-scale simulator. We aim to study: 1) what's the performance of \emph{LiveMap} as compared to existing solutions; 2) how does \emph{LiveMap} optimize offloading decision in complex automotive edge computing networks; 3) how does the data plane of \emph{LiveMap} perform over a baseline system; 4) whether \emph{LiveMap} can effectively scale under the different number of vehicles. 
In the experiments, we consider the minimum requirement of overall map coverage is 80\%, i.e., $\beta=0.8$.
The potential offloading decisions in \emph{LiveMap} are $[0,1,2,3,4]$, which correspond to offloading after the data acquisition, object detection, projection, feature extraction, and object matching module, respectively.

\begin{figure}[!t]
	\centering
	\includegraphics[width=3.48in]{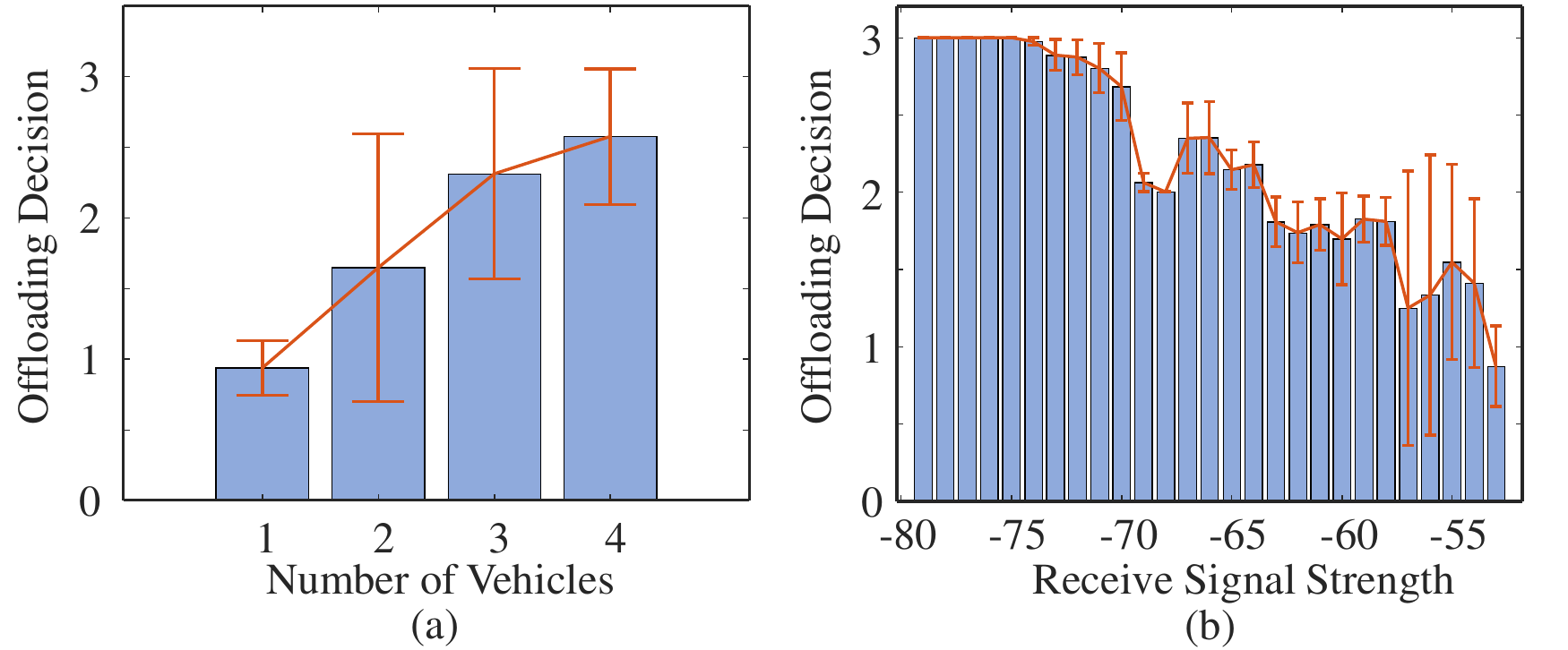}
	\caption{\small The intelligent offloading decision in \emph{LiveMap}.}
	\label{fig:intelli_decision}
\end{figure}

\begin{figure*}[!t]
	\centering
	\includegraphics[width=7.0in]{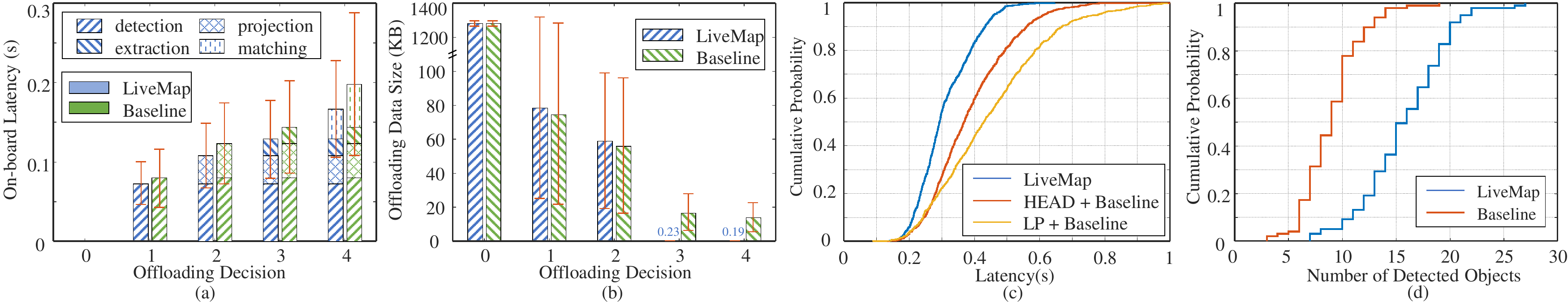}
	\vspace{-0.05in} \caption{\small The system comparison between \emph{LiveMap} and baseline.}
	\label{fig:system_comparison}
\end{figure*}


\subsection{Impact of Various Scenarios}
Fig.~\ref{fig:scenarios} shows the latency performance of different algorithms under various scenarios. 
We observe that \emph{LiveMap} achieves the lowest latency as compared to other algorithms.
In the intersection scenario, \emph{LiveMap} reduces 20.3\% average latency as compared to RM, which validates that \emph{LiveMap} can effectively schedule vehicles and intelligently determine offloading decisions, whereas model-based approach (RM) is ineffective in handling complex network system.
Meanwhile, we see \emph{LiveMap} outperforms LiveMap-Lite with 16.4\% average latency reduction, which indicates that selectively scheduling vehicles could decrease the offloading traffic in the system and thus improve the latency performance.
Furthermore, \emph{LiveMap} obtains less significant performance improvement over the other algorithms under the highway scenario, as compared to that of other scenarios.
This can be attributed to the less coverage overlap among vehicles in the highway scenario.

\subsection{Intelligent Offloading Decision}
We illustrate how \emph{LiveMap} makes the offloading decision intelligently under varying system workloads and dynamic wireless qualities.
In Fig.~\ref{fig:intelli_decision} (a), we show the statistics of offloading decisions when there are different number of vehicles in the system.
We can see that \emph{LiveMap} is prone to make larger offloading decisions, i.e., executing more modules onboard before offloading, when more vehicles are observed by the DRL agent.
In Fig.~\ref{fig:intelli_decision} (b), we show the correlations between offloading decisions and received signal strength of vehicles.
\emph{LiveMap} is likely to lower the offloading decision of a vehicle when better wireless quality, e.g., -60dBm, is observed.
Under the worst wireless quality, e.g., -80dBm, \emph{LiveMap} does not keep increasing the offloading decision, because the offloading data size after the feature extraction is similar to that of object matching.
As a result, increasing the offloading decision at such conditions will only cost more onboard execution time without reducing significant transmission delay.
These results indicate that the DRL agent can intelligently optimize the offloading decision of vehicles in \emph{LiveMap}.

\subsection{Design of Data Plane}
We show the performance of \emph{LiveMap} data plane as compared to a baseline system in terms of processing latency, offloading data size, and the number of successfully detected objects in Fig.~\ref{fig:system_comparison}.
The baseline system is implemented with tiny YOLOv3 model~\cite{redmon2018yolov3}, ORB feature extraction~\cite{rublee2011orb}, and brutal-force feature matching algorithm, and the other modules are implemented as same as \emph{LiveMap}.
In Fig.~\ref{fig:system_comparison} (a), We see that \emph{LiveMap} has lower onboard execution latency than the baseline system on different processing modules.
This is achieved by optimizing various modules in \emph{LiveMap}, e.g., lower detection time with neural network pruning in object detection, lower extraction time with autoencoder for feature extraction, and lower matching time since total object features are smaller.
Here, the object detection, including image preparation, detection network inference, and post-processing, consumes an average 72.4ms in \emph{LiveMap} and 80.2ms in the baseline system.
This is because the preparation, e.g., image reading and formatting, and the post-processing, e.g., non-maximal suppression (nms) in YOLOv3 and memory copying from GPU to CPU, account for considerable latency in the Jetson Nano embedded GPU platform.
Besides, we observe the offloading data size of \emph{LiveMap} is substantially smaller than that of the baseline system after feature extraction in Fig.~\ref{fig:system_comparison} (b).
This is attributed to the high compression ratio of autoencoder during the feature extraction, where the output latent features of an object image have only 25 values.
Here, the large variations in projection and feature extraction, i.e., offloading decision 1 and 2, come from the varying number of objects detected from RGB images. 

In Fig.~\ref{fig:system_comparison} (c), we show the cumulative probability of latency obtained by \emph{LiveMap}, the HEAD algorithm and LP algorithm in the baseline system, where \emph{LiveMap} obtains 20.1\% and 34.1\% the average latency reduction than the HEAD and LP algorithm in the baseline system, respectively.
This result validates the performance of data plane in \emph{LiveMap} is considerably better than that of the baseline system.
Furthermore, we evaluate the system performance of \emph{LiveMap} and the baseline system in Fig.~\ref{fig:system_comparison} (d), in terms of the number of successfully detected objects.
In the experiments, we consider the object is successfully detected only if its object ID is matched correctly and the estimated geo-location error is less than 1 meter. 
We can see \emph{LiveMap} substantially outperforms the baseline system with 74.9\% improvement on the average number of detected objects.
This is because the ORB feature extraction is ineffective in extracting meaningful features from small images.
As a result, the baseline system produces more matching errors, either matching to incorrect objects in the database or identifying existing objects as newly detected.
In addition, the location-aware distance function in the object matching module also contributes to the better performance of \emph{LiveMap}.
Therefore, we can conclude that the optimization of data plane in \emph{LiveMap} significantly improves the system performances.

\begin{figure}[!t]
	\centering
	\includegraphics[width=3.48in]{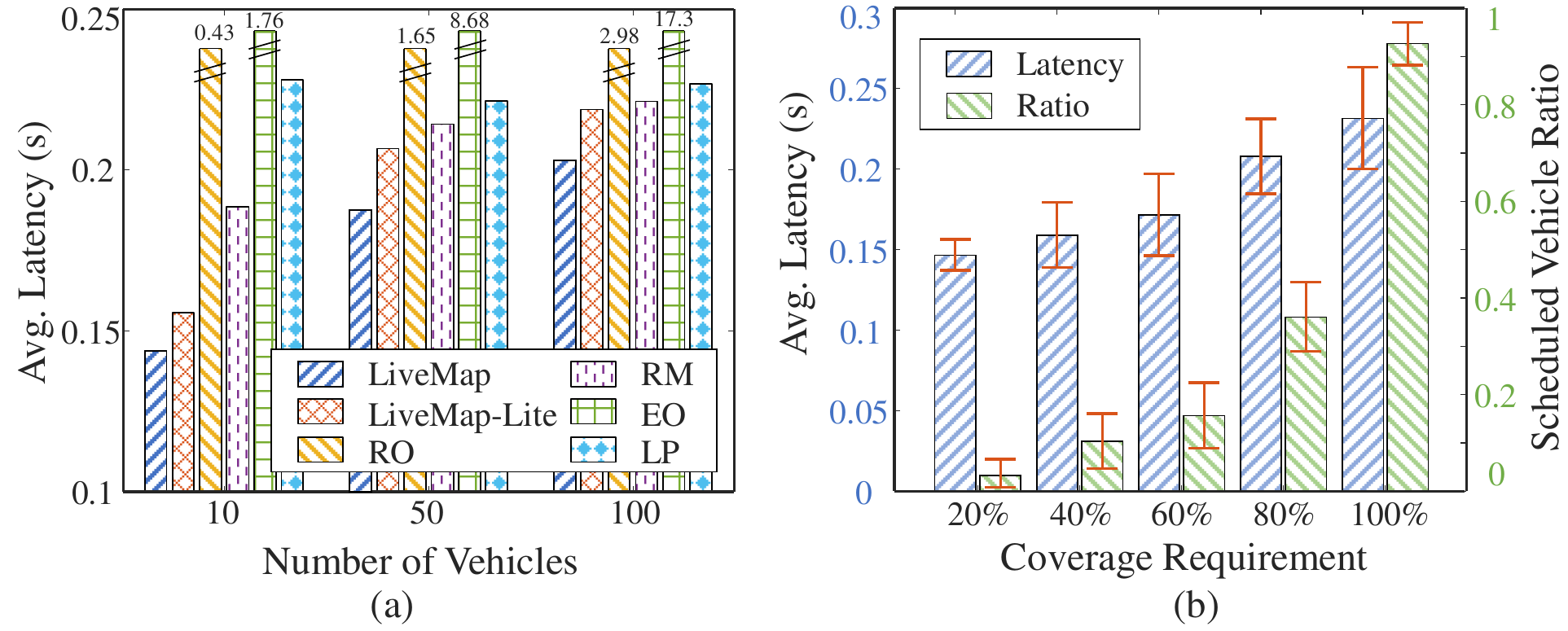}
    \caption{\small The simulated results under different number of vehicles and coverage requirements.}
	\label{fig:simulation_carnum_coverage}
\end{figure}

\subsection{Scalability in Trace-driven Simulation}
We further evaluate the performance of \emph{LiveMap} in the large scale network simulator.
In Fig.~\ref{fig:simulation_carnum_coverage} (a), we show the average latency under various algorithms with the different number of vehicles.
We observe that \emph{LiveMap} outperforms other algorithms, e.g., it reduces 10.4\% and 12.7\% latency than LiveMap-Lite and RM respectively, when there are 50 vehicles in the system.
Besides, we show the average latency and ratio of scheduled vehicles obtained by \emph{LiveMap} under different coverage requirements in Fig.~\ref{fig:simulation_carnum_coverage} (b).
We can see that \emph{LiveMap} schedules fewer vehicles with the decreasing of coverage requirement.
By sacrificing more coverage performance, e.g., from 100\% to 60\%, the scheduled vehicle ratio is decreased from 100\% to 15.6\%, and thus the average latency is reduced from 231.7ms to 159.1ms.
These results validate that \emph{LiveMap} is scalable under the different number of vehicles.


\section{Related Work}

This work relates to ML-based resource management and vehicle sensing in automotive edge computing networks.

\textbf{ML in Networking}:
Harmony~\cite{bao2019deep} exploits deep learning based scheduler to optimize the average completion time of concurrent ML tasks in cloud computing clusters. 
EdgeSlice~\cite{liu2020edgeslice} uses a decentralized DRL based approach to manage multiple network resources while meeting the service level agreement (SLA) of network slices.
DeepCast~\cite{wang2019intelligent} utilizes DRL techniques to learn the personalized quality of experience (QoE) of viewers and optimize edge servers assignment in crowdsourcing livecast.
However, these works focus on unconstrained RL problems with fixed action space, e.g., fixed number of users, while \emph{LiveMap} handles the varying number of vehicles under the requirement of map coverage.

\textbf{Vehicle Sensing}:
Augmented vehicular reality (AVR)~\cite{qiu2018avr} extends vehicular vision with V2V visual sensor data sharing to improve driving safety, where only point clouds of dynamic objects are exchanged for efficient transmission.
F-Cooper~\cite{chen2019f} fuses visual features from multiple vehicles to cooperatively perceive objects on the road, which allows the tradeoff between detection accuracy and wireless bandwidth requirements.
CarMap~\cite{ahmad2020carmap} realizes near real-time updates on the feature-represented map by excluding transient information, e.g., parked cars and pedestrians, from map processing.
However, these works focus on static information sharing, e.g., point clouds or features, where \emph{LiveMap} allows dynamic adaptive offloading, e.g., images, features or labels, for crowdsourcing vehicles in automotive edge computing networks.
%

%
%

\section{Conclusion}
In this paper, we have presented \emph{LiveMap}, a real-time dynamic map that allows efficient information sharing among connected vehicles in automotive edge computing networks.
We have developed the data plane of \emph{LiveMap} to detect, match, and track objects on the road based on the crowdsourcing data from connected vehicles in sub-second.
We have designed the control plane of \emph{LiveMap} to intelligently schedule vehicles and determine offloading decision according to the availability of networking and computation resources. 
We have demonstrated \emph{LiveMap} has better performance than existing solutions with both experimental and simulation results.

\bibliographystyle{IEEEtran}
\bibliography{ref/reference}

\end{document}